\documentclass[letterpaper,11pt]{article}

\usepackage[utf8]{inputenc}
\usepackage[T1]{fontenc}
\usepackage{cfr-lm}

\usepackage{relsize}
\usepackage{exscale}

\usepackage{bbold}

\usepackage{geometry}

\usepackage{stmaryrd}
\usepackage{cite}
\usepackage[dvips]{graphicx}
\usepackage{graphpap}
\usepackage{ifthen}
\usepackage{enumerate}
\usepackage{amssymb}
\usepackage{mathrsfs}
\usepackage[errorshow]{tracefnt}
\usepackage{fancybox}
\usepackage{amsfonts}
\usepackage{amsmath}
\usepackage{amssymb}
\usepackage{multirow}
\usepackage{array}
\usepackage{mathtools}
\usepackage{soul}
\usepackage{pict2e}
\usepackage{mathabx}
\DeclareMathAlphabet{\mathpzc}{OT1}{pzc}{m}{it}

\usepackage{lscape}
\usepackage{xypic}
\usepackage{color}
\usepackage[
color,matrix,arrow]{xy}
\usepackage{setspace}

\usepackage{enumitem}

\usepackage{graphicx}

\usepackage{keyval}
\makeatletter
\define@key{setpar}{left}[0pt]{\leftmargin=#1}
\define@key{setpar}{right}[0pt]{\rightmargin=#1}
\define@key{setpar}{both}{\leftmargin=#1\relax\rightmargin=#1}
\makeatother

\newenvironment{narrow}[1][]
  {\list{}{\setkeys{setpar}{left,right}%
     \setkeys{setpar}{#1}%
     \listparindent=\parindent
     \topsep=0pt
     \partopsep=0pt
     \parsep=\parskip}\item\relax\hspace*{\listparindent}\ignorespaces}
  {\endlist}

\newgeometry{vmargin={48mm}, hmargin={48mm}, top={48mm}, bottom={54mm}}

\makeatletter
\newcommand{\thickhline}{%
    \noalign {\ifnum 0=`}\fi \hrule height 1pt
    \futurelet \reserved@a \@xhline
}
\newcolumntype{'}{@{\hskip\tabcolsep\vrule width 1pt\hskip\tabcolsep}}
\makeatother
\newcolumntype{"}{@{\hskip\tabcolsep\vrule width 1.5pt\hskip\tabcolsep}}
\makeatother

\def\lb{{\lambda}}

\def\ie{{\it i.e.}}

\def\eg{{\it e.g.}}
\def\apriori{{\it a priori}}

\def\small#1{{\hbox{$#1$}}}
\def\fraction#1{\small{1\over#1}}
\def\fr{\fraction}
\def\Fraction#1#2{\small{#1\over#2}}
\def\Fr{\Fraction}

\def\boxit#1{\vbox{\hrule\hbox{\vrule\kern3pt
             \vbox{\kern3pt#1\kern3pt}\kern3pt\vrule}\hrule}}

\newcommand{\beq}{\begin{equation}}
\newcommand{\beqn}{\begin{equation*}}
\newcommand{\eeq}{\end{equation}}
\newcommand{\eeqn}{\end{equation*}}
\newcommand{\beqa}{\begin{eqnarray}}
\newcommand{\beqan}{\begin{eqnarray*}}
\newcommand{\eeqa}{\end{eqnarray}}
\newcommand{\eeqan}{\end{eqnarray*}}
\newcommand{\bdm}{\begin{displaymath}}
\newcommand{\edm}{\end{displaymath}}

\newcommand{\ba}{\begin{array}}
\newcommand{\ea}{\end{array}}

\newcommand\nn{\nonumber}

\newcommand\benu{\begin{enumerate}}
\newcommand\eenu{\end{enumerate}}
\newcommand\bit{\begin{itemize}}
\newcommand\eit{\end{itemize}}

\def\tr{\mathrm{tr\,}}

\def\der'{\mathfrak{der}'\,}
\def\der{\mathfrak{der}\,}
\def\str'{\mathfrak{str}'\,}
\def\str{\mathfrak{str}\,}

\def\g{\mathfrak{g}}

\def\*{\partial}

\def\lb{\bar\lambda}
\def\g{\gamma}
\def\ll{\lambda}

\RequirePackage{hyperref}

\numberwithin{equation}{section}

\begin{document}


\frenchspacing

%






\thispagestyle{empty}

\null\vspace{12mm}

\begin{center}
  {\Large \bf \sc Pure spinors in classical and quantum supergravity}

\vspace{14mm}
    
{\large
Martin Cederwall }

\vspace{14mm}

       {\footnotesize {\it Department of Physics,
         Chalmers Univ. of Technology,
 SE-412 96 Gothenburg, Sweden}}\\
\vspace{2mm}
{\footnotesize and}\\
\vspace{2mm}
       {\footnotesize{\it NORDITA,
         Hannes Alfv\'ens v\"ag 12,
 SE-106 91 Stockholm, Sweden}}

\end{center}

\vfill

\begin{quote}
  
\textbf{Abstract:} 
This is an overview of the method of pure spinor superfields, written for {\it ``Handbook of Quantum Gravity''}, eds. C. Bambi, L. Modesto and I. Shapiro. 
The main focus is on the use of the formalism in maximal supergravity on a flat background.
The basics of pure spinor superfields, and their relation to standard superspace, is reviewed.
The pure spinor superstring model of Berkovits is briefly discussed.
Consequences for divergence properties of loop diagrams in maximal supergravity are restated.
Some final remarks are made concerning desirable development of the theoretical framework.
\end{quote} 

\vfill

\hrule

\noindent{\tiny email:
  martin.cederwall@chalmers.se}

\newpage

\tableofcontents

\newpage

\section{Introduction}

Pure spinor superfield theory \cite{Cederwall:2013vba} provides a
solution to the long-standing problem of covariant quantisation of (Brink--Schwarz) superparticles 
\cite{Brink:1981nb,Casalbuoni:1976tz}
or (Green--Schwarz) superstrings 
\cite{Green:1983wt} with manifest supersymmetry, or roughly equivalently, to the problem of finding off-shell superspace formulations of maximally supersymmetric field theories, including supergravity.

Concretely, the difficulties with space-time supersymmetric particles and strings manifest themselves 
as a mixture of first and second class constraints in the same spinor.
This is the famous $\kappa$-symmetry
\cite{deAzcarraga:1982njd,Siegel:1983hh,Bengtsson:1984rw}, which is necessary for the superparticle/superstring action to describe the dynamics of a ${1\over2}$-BPS object.

In the present overview, we will not start with these superparticle or -string actions. Rather, the introduction of pure spinor variables will be motivated by the structure of the (on-shell) multiplets of maximal super-Yang--Mills theory (SYM) and supergravity (SG) in their traditional treatment on superspace.
The relation of the pure spinor formulation to the Green--Schwarz superstring is explained in ref. \cite{Berkovits:2004tw}.

The basics of the formalism is laid out in Section \ref{PureSpinorSuperfieldSection}. In Section \ref{SGSection} it is applied to supergravity, with maximal supergravity as main focus. A brief account of the pure spinor superstring theory of Berkovits is given in Section \ref{StringSection}. Quantum theory is sketched in Section \ref{QuantumSection}, and some convergence results for 
loop diagrams are restated. Finally, some remarks are made in Section \ref{RemarksSection} concerning possible refinement and development of the formalism.

The technical level of the presentation is kept at a minimum. Instead, we aim at collecting results from the sources in the reference list and present them as concisely and coherently as possible, while emphasising concepts rather than techniques.

\section{Pure spinor superfield theory\label{PureSpinorSuperfieldSection}}

Before going into a more precise derivation of pure spinor superfield formulations of specific supersymmetric models, we would like to sketch what lies at the heart of the formalism.
The supersymmetry algebra (which of course is a subalgebra of the super-Poincar\'e algebra) takes the generic form
$\{Q_\alpha,Q_\beta\}=2\gamma^a_{\alpha\beta}\*_a$. Here, $\alpha$ is some (possibly multiple) spinor index, and 
$Q_\alpha={\*\over\*\theta^\alpha}+(\gamma^a\theta)_\alpha\*_a$. Covariant fermionic derivatives 
$D_\alpha={\*\over\*\theta^\alpha}-(\gamma^a\theta)_\alpha\*_a$ satisfy $\{Q_\alpha,D_\beta\}=0$. They anticommute among themselves as $\{D_\alpha,D_\beta\}=-2\gamma^a_{\alpha\beta}\*_a$ --- flat superspace in endowed with torsion
$T_{\alpha\beta}{}^a=2\gamma^a_{\alpha\beta}$.

Suppose we introduce a bosonic spinor $\lambda^\alpha$ subject to the constraint 
\begin{align}
(\lambda\gamma^a\lambda)=0\label{PureEq}\;.
\end{align}
Then we may form a fermionic operator 
\begin{align}
Q=\lambda^\alpha D_\alpha\;,\label{FieldQ}
\end{align}
which, thanks to the constraint on $\lambda$ is nilpotent: $Q^2=0$.

It seems meaningful to consider the cohomology of $Q$, acting on functions of $x$, $\theta$ and $\lambda$.
This cohomology is guaranteed to be supersymmetric, since $Q$ anticommutes with the supersymmetry generators. It thus describes some supermultiplet.
As it turns out, any linear supermultiplet in any dimension may be obtained this way. 
In the case of on-shell multiplets the virtue of the formalism is even greater, since it seems to offer a natural way to an off-shell formulation by relaxation of the linear ``equation of motion'' $Q\Psi=0$.
The correspondence will be made more precise below, first for $D=10$ super-Yang--Mills theory and later for $D=11$ supergravity.

A word of caution: We will refer to a spinor $\lambda$ subject to the constraint \eqref{PureEq} as a ``pure spinor''.
This is a slight misuse of the mathematical terminology. A pure spinor, in the sense of Cartan \cite{Cartan:1938jua}, is a chiral spinor in even dimension $D=2n$, constrained to lie in the minimal $Spin(2n)$ orbit of the spinor module. This implies that, if the Dynkin label of the spinor module in question is $(0\ldots01)$, monomials of degree of homogeneity $p$ in $\lambda$ belong to the single module
$(0\ldots0p)$. The concept of a pure spinor is not defined in other cases, neither for odd dimensions or for extended supersymmetry. In certain cases, our constrained spinors coincide with Cartan pure spinors. This happens notably in $D=10$. There, the spinor bilinears are a vector $(10000)$ and a self-dual 5-form $(00002)$, and the constraint in the vector immediately puts $\lambda$ in the minimal orbit. In other situations, for example $D=11$ which we will encounter later, where the symmetric spinor bilinears are a vector, a $2$-form and a $5$-form, the vector constraints puts $\lambda$ in (a completion of) an intermediate orbit, which is not the minimal one.

\subsection{From superspace to pure spinor superspace}

Although we ultimately aim at addressing supergravity, the introduction of pure spinor superspace is much simpler in the setting
of super-Yang--Mills theory \cite{Brink:1976bc}, first treated in superspace in ref. \cite{Siegel:1978yi}.

Flat $(10|16)$-dimensional superspace, appropriate for $D=10$ super-Yang--Mills, has coordinates $Z^M=(x^m,\theta^\mu)$. There is no metric on superspace, but a super-vielbein $E_M{}^A$. The Lorentz frame indices $A=(a,\alpha)$ consist of a Lorentz vector and a chiral spinor. The non-vanishing superspace torsion is 
\begin{align}
T_{\alpha\beta}{}^a=2\gamma^a_{\alpha\beta}\;,
\end{align}
where the components are converted to Lorentz frame using the inverse vielbein: 
$T_{BC}{}^A=(E^{-1})_C{}^N(E^{-1})_B{}^MT_{MN}{}^A$.

Let us now recollect some known facts about $D=10$ super-Yang--Mills.
A connection on superspace, taking values in the adjoint of some gauge group, is written
$A=dZ^MA_M=dZ^ME_M{}^AA_A$.
There is \apriori\ two superfields, $A_a(x,\theta)$ (bosonic, dimension $1$) and $A_\alpha(x,\theta)$ 
(fermionic, dimension $\fr2$)\footnote{As is standard, dimension is in terms of powers of inverse length.}.
The field strength is $F=dA+A\wedge A$, and due to the presence of torsion we have
\begin{align}
F_{\alpha\beta}=2D_{(\alpha}A_{\beta)}+2A_{(\alpha}A_{\beta)}+2\gamma^a_{\alpha\beta}A_a\;.\label{F0Eq}
\end{align}
The symmetric product of two spinors can be decomposed into a vector 
$F_a=\fr{16}\gamma_a^{\alpha\beta}F_{\alpha\beta}$ 
and a self-dual $5$-form 
$F_{abcde}=\fr{2\cdot5!\cdot16}\gamma_{abcde}^{\alpha\beta}F_{\alpha\beta}$.
Obviously, from eq. \eqref{F0Eq}, setting $F_a=0$ expresses $A_a$ in terms of $A_\alpha$, leaving only the latter as an independent superfield. This goes under the name of ``conventional constraint''. Note that it is natural, since $F_{\alpha\beta}$ has dimension $1$, and there are no physical and gauge-covariant fields of this dimension in the supermultiplet we want to derive
(the spinor $\chi^\alpha$ has dimension $\Fr32$ and the field strength $F_{ab}$ dimension $2$). 
For this reason, it is tempting to also set the $5$-form part $F_{abcde}$ to zero. 
However, doing this turns out to put the theory on shell. It indeed describes the on-shell $D=10$ super-Yang--Mills multiplet.
The constraint in question is physical, rather than conventional.
Details can be found \eg\ in refs. \cite{Nilsson:1985cm,Cederwall:2001bt}.
It is of course also well known that the supersymmetry transformations of the component fields work (``close'', modulo gauge transformations) only when the equations of motion are satisfied. There is no set of auxiliary fields that remedies this.
 
This observation prompted Nilsson \cite{Nilsson:1985cm} to first draw the (correct) conclusion that in order to go off-shell one needs to relax the $5$-form part of $F_{\alpha\beta}=0$. And this is what pure spinor superfield theory naturally does, as we will see. Indeed, the equations of motion in the pure spinor superfield description of $D=10$ super-Yang--Mills theory will be 
$\gamma_{(5)}^{\alpha\beta}F_{\alpha\beta}=0$.

This structure was found when searching for deformations of the equations of motion for maximally supersymmetric super-Yang--Mills \cite{Cederwall:2001bt,Cederwall:2001td,Cederwall:2001dx}.
Introduce a bosonic spinor $\lambda^\alpha$, subject to the constraint $(\lambda\gamma^a\lambda)=0$. 
A function $\Psi$ of $x$, $\theta$ and $\lambda$, expanded in powers in $\lambda$, is\footnote{There is no other way of dealing with the $\lambda$ dependence, since no scalar is encountered at any power of $\lambda$.}
\begin{align}
\Psi(x,\theta,\lambda)=C(x,\theta)+\lambda^\alpha A_\alpha(x,\theta)
+\fr2\lambda^\alpha\lambda^\beta B_{\alpha\beta}(x,\theta)+\ldots
\end{align}
Now, acting with the ``BRST operator'' $Q$ gives
\begin{align}
Q\Psi=\lambda^\alpha D_\alpha C+\lambda^\alpha\lambda^\beta D_\alpha A_\beta+\ldots
\end{align}
The linearised equations of motion are encoded as $\Psi$ being $Q$-closed, and gauge transformations correspond to $Q$-exact functions.
It also immediately follows that, for the specific choice $\Psi=\lambda^\alpha A_\alpha$,
\begin{align}
Q\Psi+\Psi^2=\lambda^\alpha\lambda^\beta F_{\alpha\beta}\;.
\end{align}
The full non-linear equations of motion are encoded as 
\begin{align}
Q\Psi+\Psi^2=0\label{SYMEOM}\;.
\end{align}

One should think of $\lambda$ as a ghost variable, which explains it being bosonic, although it is a spinor.
Then $\Psi$ should also be assigned ghost number $1$, so that $A_\alpha$ has ghost number $0$. 
The above shows that the cohomology of $Q$ in the ghost number $0$ sector is precisely the linear super-Yang--Mills multiplet. One should also make sure that there is no essential cohomology in other ghost numbers (powers of $\lambda$). This can be done as follows.

In order to investigate the cohomology, we will do it in two steps: first we find the zero-mode (\ie, $x$-independent) cohomology. It will correspond to fields in in a component formulation. Then, in the second step, these fields will, in the full cohomology, be related by differential operators constructed from ${\*\over\*x}$. The procedure is not presented as a mathematical proof here; a fuller account can be found in refs. \cite{Cederwall:2013vba,Eager:2021wpi,CederwallJonssonPalmkvistSaberi}.

The zero-mode cohomology of $Q$ is the cohomology of $\lambda^\alpha{\*\over\*\theta^\alpha}$. Had $\lambda$ been unconstrained, the only cohomology would have been the constant one. Now, when $\lambda$ is constrained, the problem is algebraic, and the result is reflected in the partition function of $\lambda$. Encode the power of $\lambda$ in a variable $t$. Then the partition function, taking values in the representation ring, is
\begin{align}
Z(t)=(00000)\oplus(00001)t\oplus(00002)t^2\oplus(00003)t^3+\ldots\label{LambdaPartition1}
\end{align}
It is straightforward to factor out the dependence of an unconstrained spinor, which we write as
\begin{align}
(1-t)^{-(00001)}=(00000)\oplus(00001)t\oplus\vee^2(00001)t^2\oplus\vee^3(00001)t^3+\ldots
\end{align}
($\vee^p$ is the $p$-fold totally symmetric product). We then obtain
\begin{align}
&Z(t)=(1-t)^{-(00001)}\\
&\otimes\left[
(00000)\ominus(10000)t^2\oplus(00010)t^3\ominus(00001)t^5\oplus(10000)t^6\ominus(00000)t^8
\right]\;.\nn\label{LambdaPartition2}
\end{align}
Each term inside the square brackets represents a component field. They are, in order of appearance,
the ghost $c$, the fields in the physical multiplet $A_a$ and $\chi^\alpha$, the antifields $\chi^{\star}_\alpha$ and $A^{\star a}$, and the ghost antifield $c^\star$. Each of them appears in the zero-mode cohomology as some function of $\theta$ and $\lambda$, for example the physical gauge field appears as $(\lambda\gamma^a\theta)A_a$.

Going back to the full cohomology of $Q$, it will relate the component fields, now $x$-dependent, with differential operators.
The proper mathematical tool for this procedure is that of homotopy transfer, see ref. \cite{Eager:2021wpi}.
It is straightforward to show that the action of $Q$ is indeed that of the BRST operator of the component fields and antifields of the super-Yang--Mills theory. The cohomology consists precisely of the linearised physical (on-shell) fields, and a ghost zero-mode.

It looks tempting to try to derive eq. \eqref{SYMEOM} from a Chern--Simons like action. This can indeed be done, leading to the appropriate off-shell formulation, but requires the machinery of the following subsection.
It is also clear from the nature of the cohomology that such an action should be regarded as a Batalin--Vilkovisky (BV) action
\cite{Batalin:1981jr}, containing ghosts, fields and their antifields.

\subsection{Non-minimal variables, integration, BV actions}

In order to write an action that reproduces the equations of motion of the previous subsection, 
one needs an integration over the pure spinor $\lambda$. In addition, it should (when one also includes integration over $\theta$) pick up the top zero-mode cohomology, \ie, the top component of the component super-Yang--Mills BRST complex, corresponding to the ghost antifield. This cohomology sits at $\lambda^3\theta^5$. One is in the seemingly problematic situation of needing a residue-like measure, in the sense of picking a certain component, while on the other hand having a series expansion that contains only positive powers of $\lambda$. Such a measure is clearly degenerate, and not useful.

This problem was solved in ref.
\cite{Berkovits:2005bt}, using what is known as a non-minimal set of variables.
In addition to the pure spinor $\lambda^\alpha$, one introduces a conjugate pure spinor $\bar\lambda_\alpha$. In order not to disturb the cohomology, an equal number of additional fermions $r_\alpha$, which are pure with respect to $\bar\lambda$: 
$(\bar\lambda\gamma^a r)=0$. We identify $r_\alpha$ as $d\bar\lambda_\alpha$, and products of $r$'s with wedge product of $d\bar\lambda$'s. Then, the modified non-minimal BRST operator
\begin{align}
Q=(\lambda D)+\bar\*\;,
\end{align}
where $\bar\*=d\bar\lambda_\alpha{\*\over\*\bar\lambda_\alpha}$ is the Dolbeault operator,
has the same cohomology as the minimal one previously considered.

The pure spinor space is a (non-compact) Calabi--Yau space \cite{Cederwall:2011yp}. It possesses a holomorphic top form, in this case an $11$-form $\Omega$. The schematic form of this Calabi--Yau form is
\begin{align}
\Omega\sim\lambda^{-3}(d\lambda)^{11}\;.
\end{align}
For detailed expressions, see refs. \cite{Berkovits:2005bt,Cederwall:2011yp,Cederwall:2013vba}.

Remember that the pure spinor field $\Psi$ now depends on $x$, $\theta$, $\lambda$, $\bar\lambda$ and $d\bar\lambda$. The last dependence is seen as $\Psi$ being an antiholomorphic cochain. One may try an integration measure
\begin{align}
\int [dZ] f=\int d^{10}x\int d^{16}\theta\int \Omega\wedge f\;,
\end{align}
where the last integral is over the pure spinor Calabi--Yau space.
This measure is non-degenerate, and carries ghost number $-3$ as desired, due to the $\lambda^{-3}$ in $\Omega$. However, the cohomologies we encountered have representatives which are $0$-forms, so any pair of such functions seems to have vanishing scalar product. On the other hand, the pure spinor space is a non-compact c\^one, so integrals na\"ively diverge at large radius.
This ``$0\times\infty$'' structure can be regularised \cite{Berkovits:2005bt} to yield finite results.
The trick is to observe that the behaviour on pure spinor space is topological, and to insert a $Q$-invariant regularisation
\cite{Marnelius:1990eq,Berkovits:2005bt}
$e^{-t\{Q,\chi\}}$ for some fermion $\chi$. Such a regulator will give $t$-independent results. 
If one chooses $\chi=\theta^\alpha\bar\lambda_\alpha$, one gets a factor
\begin{align}
e^{-t\{Q,\chi\}}=e^{-t((\lambda\bar\lambda)+(\theta d\bar\lambda))}\;.\label{SimpleRegEq}
\end{align}
The first factor makes integrals convergent at large radius. The second one contains terms up to $\theta^{11}(d\bar\lambda)^{11}$. When integrated with a $0$-form, it will pick up a component at $\theta^5$.
This regulated measure is exactly what is needed. 
We can think of it as an operator that localises the integral to the vicinity of the tip $\lambda=0$ of the pure spinor c\^one.
Alternatively, the basis for cohomology can be chosen to include such factors, and then no regularisation of the measure is necessary.

Now, a Chern--Simons-like BV action for $D=10$ can be written as \cite{Berkovits:2001rb,Movshev:2003ib}
\begin{align}
S=\int[dZ]\tr(\fr2\Psi Q\Psi+\fr3\Psi^3)\;.
\end{align}
Note that the action only contains a cubic interaction term, while the component $F^2$ contains quartic interactions.
A component action can be derived by homotopy transfer \cite{Eager:2021wpi}, or put more mundanely, the higher order interactions arise from repeated use of the equations of motion. See also ref. \cite{Berkovits:2018gbq}.
This property, that the supersymmetric action is of lower order than the component action, becomes even more pronounced when we turn to supergravity in the following Section.

The superfield $\Psi$ is self-conjugate with respect to the BV anti-bracket:
\begin{align}
(A,B)=\int A{\overleftarrow\delta\over\delta\Psi(Z)}[dZ]{\overrightarrow\delta\over\delta\Psi(Z)}B\;.
\end{align}
Then it is straightforward to show that the classical master equation $(S,S)=0$ is satisfied.

\subsection{Other models}

Any supermultiplet can be derived as the cohomology in a pure spinor superfield.
In many cases, the zero-mode cohomology is such that the corresponding fields define a component BV complex, and an integration can be defined. Situations where this does not happen is \eg\ when self-dual tensors are contained in the supermultiplet, such as the $N=(2,0)$ multiplet in $D=6$ or type IIB supergravity. 
If the multiplet has an off-shell formulation with some auxiliary fields, 
this off-shell multiplet is found as the cohomolgy of $Q$ \cite{Cederwall:2008zv,Cederwall:2017ruu}. 
In such cases, the pure superspinor complex only contains ghosts and fields, and anti-fields to these are found in a conjugate pure spinor superfield \cite{Cederwall:2017ruu}.
In many cases one needs to use pure spinor superfields in non-trivial modules
\cite{Cederwall:2011vy,Cederwall:2008vd,Cederwall:2008xu,Cederwall:2017cez,Cederwall:2020dui,Cederwall:2021ejp} (see for example the field $\Phi^a$ in Section
\ref{Geometry4FormSection}). In the language of ref. \cite{Eager:2021wpi}, they belong to sections of some sheaf over the pure spinor space.

Higher derivative deformations of supersymmetric models, for example supersymmetric Born--Infeld theory may be given simple (polynomial) actions in pure spinor superfield theory \cite{Cederwall:2011vy,Chang:2014nwa}.

Supergravity, in particular in $11$ dimensions, will be addressed in the following Section.

\subsection{Pure spinor partition functions and superalgebras}

Given the usefulness of pure spinors for the description of supermultiplets in general, it seems meaningful to pursue a deeper mathematical investigation of the algebraic properties of pure spinor space itself. Functions on pure spinor space can be thought of 
algebraically as power series in $\lambda$, modulo the ideal generated by $(\lambda\gamma^a\lambda)$.
The partition function of eqs. \eqref{LambdaPartition1}, \eqref{LambdaPartition2} can indeed be understood as the partition function of the on-shell super-Yang--Mills multiplet by factoring out also a level $2$ vector:
\begin{align}
Z(t)=&(1-t)^{-(00001)}\otimes(1-t^2)^{(10000)}\\
&\otimes
\left[(00000)\oplus\bigoplus\limits_{n=0}^\infty\bigl((n0010)t^{3+2n}\ominus(n1000)t^{4+2n}\bigr)\right]\;.\nn
\end{align}
The factor in the square represents the ghost zero-mode and the on-shell $n$'th derivative of the fermion and the field strength. 
This is for the $D=10$ super-Yang--Mills example. Similar statements hold for any multiplet.
The first two factors are cancelled by the partition functions for functions of $\theta$ and $x$. In this way it becomes clear that 
the pure spinor entirely encodes a full supermultiplet.

The investigation of the partition function of a pure spinor through the ghost structure associated to the bilinear constraint was initiated by Chesterman \cite{Chesterman:2002ey}, and refined by Berkovits and Nekrasov
\cite{Berkovits:2005hy}.
Consider the BRST operator $q$ for the pure spinor constraint.
It will (generically) involve an infinite number of ghosts due to the infinite reducibility of the constraint.
One may think of $q$ as the coalgebra differential of a superalgebra, and the content of the algebra as a vector space may be deduced from a continued factorisation of the partition function
\begin{align}
Z(t)=\prod\limits_{n=1}^\infty(1-t^n)^{R_n}\;.
\end{align}
One has to remember that statistics are switched and modules are conjugated when going from the ghosts (coalgebra elements) to the superalgebra. 
The superalgebra in question,
which is our definition of the Koszul dual to the functions of a pure spinor, will always be some deformation of the direct sum of the supersymmetry algebra (levels $1$ and $2$) and the freely generated algebra on the supermultiplet (levels $n\geq3$)
\cite{CederwallJonssonPalmkvistSaberi}.
The Koszul duality can be interpreted as a denominator formula for the superalgebra.
In cases where the superfield is not a scalar, this is expected to generalise to character formulas for representations of the superalgebra.

When the constraint puts $\lambda$ in a minimal orbit, the superalgebra is a Lie superalgebra, more precisely a Borcherds superalgebra \cite{Cederwall:2015oua}.
For the particular case of $D=10$ super-Yang--Mills theory, the corresponding Borcherds superalgebra in fact exactly encodes the structure of {\it interacting} super-Yang--Mills theory
\cite{Movshev:2003ib}. This is a quite amazing and unexpected result, since all that is described by the cohomology is 
the linear multiplet.
It is not yet clear what the corresponding statement is for other theories, \eg\ $D=11$ supergravity, but partial results exist
\cite{Jonsson:2021lwa,CederwallJonssonPalmkvistSaberi}. There, the superalgebra is not a Lie superalgebra, but and $L_\infty$ algebra involving (at least) a $3$-bracket and a $4$-bracket.

\section{$D=11$ supergravity\label{SGSection}}

We will not turn to $D=11$ supergravity \cite{Cremmer:1978km}.
The pure spinor superfield formulation of this model can be derived from its traditional superspace
\cite{Wess:1977fn,Brink:1978iv} formulation \cite{Brink:1980az,Cremmer:1980ru} in much the same way as the super-Yang-Mills theory was in Section \ref{PureSpinorSuperfieldSection}, however with some additional ingredients.

Recall the component field content of the $D=11$ supergravity multiplet: the metric $g_{mn}$, a $3$-form $C$ with a $4$-form field strength $H=dC$, and the gravitino field $\chi_m{}^\alpha$ with field strength $\psi_{mn}{}^\alpha$.
An essential feature, that was used as a guideline for the construction of the supersymmetric action, is that supersymmetry demands the presence of a Chern--Simons term $\int C\wedge H\wedge H$.

\begin{figure}

\begin{center}

\includegraphics[width=.5\linewidth]{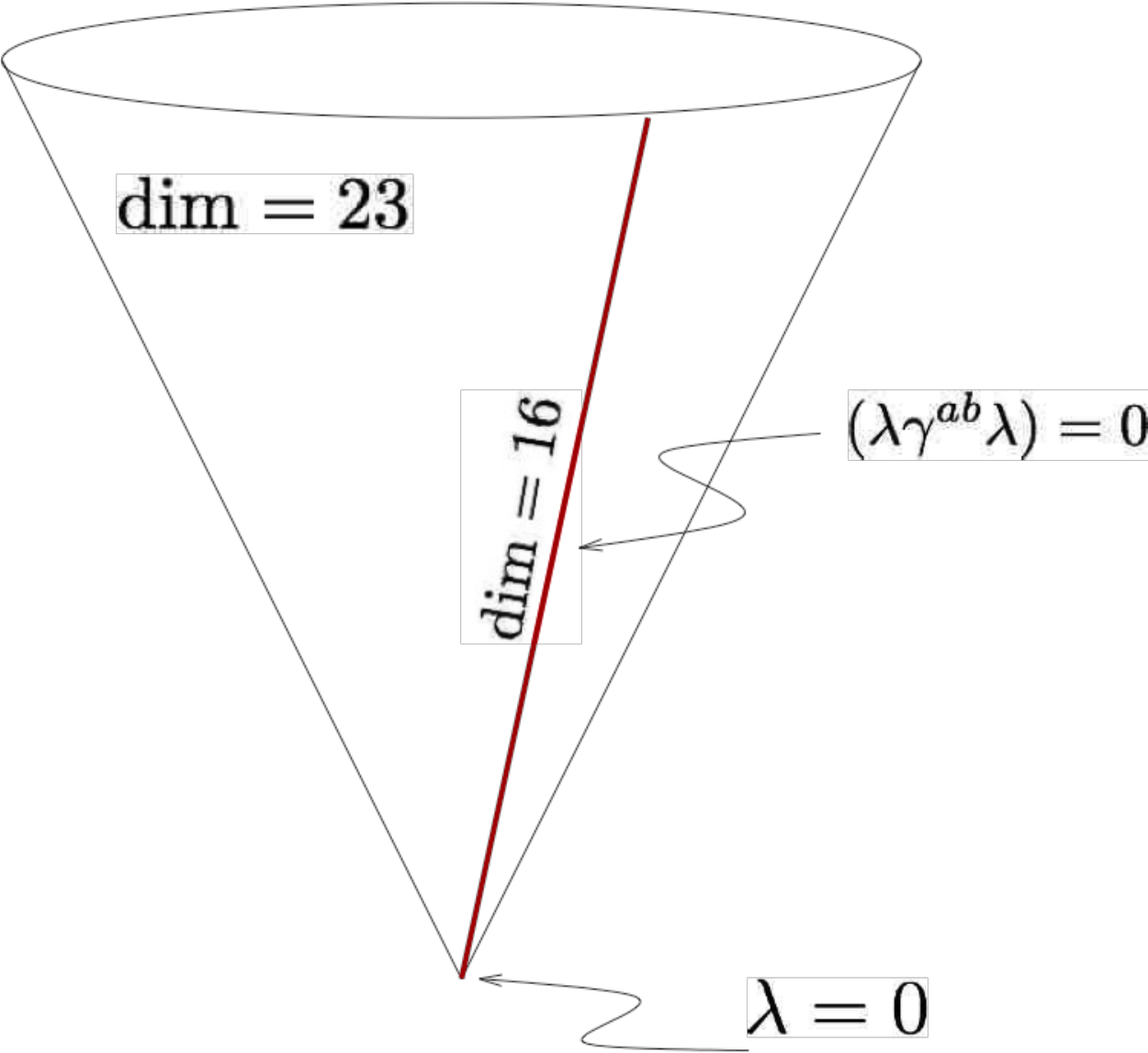}

\begin{narrow}[both=1cm]
\caption{A sketch of the space of $D=11$ pure spinors.
\label{11coneFig}}  
\end{narrow}

\end{center}

\end{figure}

In what follows, we will use $11$-dimensional ``pure spinors'' $\lambda^\alpha$. A Dirac spinor in $D=11$ has $32$ components.
The symmetric spinor bilinears consist of a vector, a 2-form and a 5-form, constructed with
$\gamma^a_{\alpha\beta}$,  $\gamma^{ab}_{\alpha\beta}$ and $\gamma^{abcde}_{\alpha\beta}$.  
A spinor subject to $(\lambda\gamma^a\lambda)=0$ is not necessarily in a minimal orbit, which would require also 
$(\lambda\gamma^{ab}\lambda)=0$. Rather, the pure spinor space consists of a ``generic'', $23$-dimensional part, complemented by the $16$-dimensional minimal orbit, which is a singular subspace, and the zero orbit, the tip of the c\^one. The space is sketched in Figure \ref{11coneFig}.

\subsection{Geometry vs. $4$-form\label{Geometry4FormSection}}

There are two different versions of superfields that can describe the on-shell (linearised) supergravity multiplet.
One relies on the standard description of superspace geometry, where one introduces a dynamical super-vielbein
$E_M{}^A$. Then, conventional constraints are used to eliminate all components except the lowest-dimensional ones,
$E_\mu{}^a$ in a controlled and covariant way \cite{Cederwall:2000ye,Cederwall:2000fj,Cederwall:2004cg}. 
Note that physical fields then are described by $1$-forms (in fermionic indices), but with an extra index $a$, while superdiffeomorphisms can be thought of as sitting in a superfield $\xi^a$ with the bosonic diffeomorphism parameters as leading components. The situation reminds of the treatment of super-Yang--Mills theory in the previous Section, although all fields have an extra index $a$. It can indeed be verified that the linearised (around Minkowski space) multiplet is described by the cohomology of $Q=(\lambda D)$ on a field $\Phi^a(x,\theta,\lambda)$. 
The field is in addition required to have a ``shift symmetry''  
\cite{Cederwall:2009ez,Cederwall:2010tn,Cederwall:2011vy}
$\Phi^a\simeq\Phi^a+(\lambda\gamma^a\varrho)$ for an arbitrary parameter 
$\varrho^\alpha(x,\theta,\lambda)$. (The shift symmetry ties together the index structure with the cohomology, and is also directly responsible for the presence of the fermionic diffeomorphism ghosts in the zero-mode cohomology.)

The above is one way to relate the on-shell linearised supergravity multiplet to pure spinor superfield cohomology. Since it is geometrical, it carries no information about the gauge symmetry of the $3$-form $C$, which indeed only appears through its field strength $H$ in the dimension $1$ torsion. 
The other way of reproducing the linearised multiplet is through a scalar field. The full ghost system for the $C$ field contains a ghost, a ghost-for-ghost and a ghost-for-ghost-for-ghost. The latter is a fermionic $0$-form. We can think of it as the $\theta$- and $\lambda$-independent zero-mode cohomology of a pure spinor superfield $\Psi$ of ghost number $3$ and dimension $-3$.
A careful calculation of the zero-mode cohomology gives at hand that it indeed contains the mentioned ghosts, together with the super-diffeomorphism ghosts (at $\lambda^2$), the physical fields (at $\lambda^3$) and all corresponding antifields.
We refer to refs. \cite{Cederwall:2009ez,Cederwall:2010tn,Cederwall:2013vba} for the detailed calculations.

Now we are in a situation where the traditional supergeometric approach gives the full non-linear equations of motion, but does not account for the full ghost structure. The scalar field $\Psi$, on the other hand, is more fundamental in that it contain all ghosts, and also the potential $C$, but it is \apriori\ unclear how to go beyond the linearised level.
Importantly, in order to write down an action containing the Chern--Simons term, $\Psi$ is needed.

\subsection{BV action}

Before giving the form of the full non-linear action, we need to understand integration, regularisation etc. in a way analogous to the $10$-dimensional case. We will refrain from detailed expressions.
The pure spinor space is $23$-dimensional. We again introduce non-minimal variables $\bar\lambda$ and $d\bar\lambda$, and 
include the Dolbeault operator $\bar\*$ in $Q$. The top cohomology of the third order ghost antifield now sits at
$\lambda^7\theta^9$. A measure based on this cohomology has the correct ghost number $-7$ for an action with $\int\Psi Q\Psi$, where $\Psi$ carries ghost number $3$. The pure spinor space is again Calabi--Yau, with $\Omega\sim\lambda^{-7}(d\lambda)^{23}$. 
A completely analogous regularisation will contain $\theta^{23}(d\bar\lambda)^{23}$, so the $\theta$ integration will effectively pick out a term with $\theta^9$ ($9=32-23$), as desired.
A linearised action
\begin{align}
S_2=\fr2\int[dZ]\Psi Q\Psi
\end{align}
reproduces the on-shell multiplet correctly.

How are interactions constructed as additional terms in a BV action?
One starting point may be to look at the Chern--Simons term $\int C\wedge H\wedge H$. 
It must contain at least one field $\Psi$, but the remaining factors can in principle be formed from $\Phi^a$, containing the field strength $H$. The concrete task now becomes to find an expression for $\Phi^a$ in terms of $\Psi$, such that cohomology maps to cohomology. This means that one needs to find a bosonic operator $R^a$ of ghost number $-2$ and dimension $2$ which commutes with $Q$ modulo terms of the type $(\lambda\gamma^a\varrho)$.
The procedure is similar to that of finding a $b$ operator, used in gauge fixing (see Section \ref{QuantumSection}).
Such an operator was constructed in ref. \cite{Cederwall:2009ez} using non-minimal variables. It takes a somewhat complicated form, beginning as 
\begin{align}
R^a=((\lambda\gamma_{cd}\lambda)(\bar\lambda\gamma^{cd}\bar\lambda))^{-1}(\bar\lambda\gamma^{ab}\bar\lambda)\*_b
+\ldots
\end{align}
One can then use $\Psi$ as the fundamental field, and write
$\Phi^a=R^a\Psi$.
A term 
\begin{align}
S_3=\fr6\int[dZ](\lambda\gamma_{ab}\lambda)\Psi R^a\Psi R^b\Psi
\end{align}
is then guaranteed to fulfil $(S_2,S_3)=0$, \ie, work as a linear deformation of $S_2$  \cite{Cederwall:2009ez}.
Note that the factor $(\lambda\gamma_{ab}\lambda)$ serves several purposes: 
It contracts the indices on a pair of fermionic fields.
It ensures the correct ghost number and dimensions of the interaction term.
And finally, it ensures the invariance under the shift symmetry, thanks to the Fierz identity
$(\gamma^b\lambda)_\alpha(\lambda\gamma_{ab}\lambda)=0$, which holds for pure spinors.
It can be verified, using explicit expressions for the cohomologies, that the Chern--Simons term is correctly reproduced by $S_3$.

In order to construct a complete action, the master equation $(S,S)=0$ must be checked, not only to linear order in $S_3$ as above. It turns out \cite{Cederwall:2010tn} that only a minor modification is needed: a $4$-point coupling which is almost of the same form as $S_3$. It relies on yet another operator, $T$, of ghost number $-3$ and dimension $3$. The field $T\Psi$ then carries ghost number and dimension $0$, and its ghost number $0$ part can be thought of as containing the trace of the linearised gravity field.
Then, the action
\begin{align}
S=\int[dZ]\left(\fr2\Psi Q\Psi+\fr6(\lambda\gamma_{ab}\lambda)(1-\Fr32T\Psi)\Psi R^a\Psi R^b\Psi\right)
\label{FullSGActionEQ}
\end{align}
turns our to satisfy $(S,S)=0$ to all orders,
It is striking, but ideal from the point of view of perturbative calculations, that a model containing gravity becomes polynomial around Minkowski space. A detailed understanding, \eg\ through homotopy transfer, of how the non-polynomial nature of geometry around Minkowski space arises, is still lacking. Neither does the construction offer any direct clues concerning how to proceed to other backgrounds. Some remarks concering these issues are given in the concluding Section.

\subsection{Twisting}

Pure spinor superfields provide a good framework for twisting supersymmetric theories, and to find all possible twistings
\cite{Eager:2018dsx,Saberi:2021weg}. This is because any point in the space of spinors obeying
$(\lambda\gamma^a\lambda)=0$ provides a nilpotent operator $\lambda^\alpha D_\alpha$. (Note that here $\lambda$ is not a variable, but takes some specific value.) 
The list of possible twistings can be read off from the stratification of pure spinor space in different orbits under the Lorentz group, forming subspaces of pure spinor space.

In supergravity, supersymmetry is local, and twisting is performed by giving an expectation value to a superdiffeomorphism ghost
\cite{Costello:2016mgj}. A treatment in the pure spinor formalism is favourable, since these ghosts are naturally present.
Among other theories, the twistings of $D=11$ supergravity has been thus examined
\cite{Raghavendran:2021qbh,Saberi:2021weg,Eager:2021ufo}.
The minimal twist leads to the $SL(5)$ supersymmetric model of ref. 
\cite{Cederwall:2021ejp}.

\section{Superstrings\label{StringSection}}

The covariant quantisation of space-time supersymmetric string theory remained elusive for a long time, until
Berkovits constructed the pure spinor superstring 
\cite{Berkovits:2000fe,Berkovits:2000ph,Berkovits:2000nn,Berkovits:2017ldz}.
The variables used are the same as displayed above for $D=10$ super-Yang--Mills theory. 
In both the left- and right-moving sectors of the world sheet, one introduces in addition to the superspace coordinates $X$ (self-conjugate) and $\theta$, with its conjugate $p$, a pure spinor $\lambda$ and its conjugate $\omega$.
The variable $\lambda$ has the same chirality as $\theta$. In type IIA superstring theory this chirality is opposite for left- and right-movers, and in type IIB the same.

The left-moving BRST operator reads 
\begin{align}
Q=\oint\lambda^\alpha d_\alpha\;,\label{StringQ}
\end{align}
where
\begin{align}
d_\alpha=p_\alpha+\*X^a(\gamma_a\theta)_\alpha+\fr8(\gamma^a\theta)_\alpha(\theta\gamma_a\*\theta)\;,
\end{align}
with the operator product expansion
\begin{align}
d_\alpha(z)d_\beta(\zeta)={1\over z-\zeta}\gamma_{a\alpha\beta}\Pi^a+\hbox{(regular)}\;,
\end{align}
where $\Pi^a=\*X^a-(\theta\gamma^a\*\theta)$ is the momentum conjugate to $X$ in the Green--Schwarz superstring.
This implies $Q^2=0$. Again, it can of course be extended with non-minimal variables.

Notably the list of fields above is complete, including ghosts. There is no Virasoro ghost pair $(b,c)$ (and, unlike the Neveu--Schwarz--Ramond superstring, no super-Virasoro ghosts $(\beta,\gamma)$). 
This of course also happens for the superparticle.
All ``coordinates'' are world-sheet scalars. 
The cancellation of the conformal anomaly requires no Virasoro ghost, but simply reads
$c=10-2\cdot 16+2\cdot 11=0$. This may seem as a simplification, but also has its price in making \eg\ gauge fixing more complicated, see Section \ref{QuantumSection}.

Integration over pure spinor variables follows the same principles as for the super-Yang--Mills theory.

\section{Quantum theory\label{QuantumSection}}

The procedures sketched in this Section
focus on principles and qualitative results.
The issue of gauge fixing and the $b$ operator is discussed in somewhat more detail, since this is one of the
points where the formalism becomes complicated and simplifications are desired.
The physical fields are ``hidden'' within a structure which exhibits many qualitatively simple features. Their extraction from that structure is more complicated \cite{Eager:2021wpi,Berkovits:2018gbq}. If one wants to use the formalism to derive precise quantitative results, much work is involved (see refs. below). 

\subsection{Gauge fixing\label{GaugeFixingSection}}

The non-minimal variables open for a possibility to construct operators with negative ghost number. 
The so-called ``$b$ ghost'', or $b$ operator, is the standard example (see also the negative ghost number operators of Section
\ref{SGSection}).
It is called $b$ because it assumes the r\^ole of the conjugate to the ghost $c$ for world-line reparametrisations or world-sheet conformal transformations (see the cancellation of the conformal anomaly in Section \ref{StringSection}). In pure spinor superfield theory it is a composite operator. This is because ``$p^2=0$'' (in a superparticle action) is only a derived linearised equation of motion, a consequence (after gauge fixing) of $Q\Phi=0$, not a constraint associated with a world-line symmetry. 

In order to perform perturbative quantum calculations gauge fixing is necessary. The ``kinetic operator'' $Q$ is of course not invertible. If one can find an operator $b$ such that $\{Q,b\}=\Box$ and chooses the Siegel gauge \cite{Siegel:1984ogw}
\begin{align}
b\Psi=0\;,
\end{align}
 the propagator $G$ can be written as 
\begin{align}
G={b\over\Box}\;.
\end{align}
In the following we will write the field theory $b$ operator. The one for string theory is very similar (in the same way as the $Q$'s of eqs. \eqref{FieldQ} and \eqref{StringQ} are), just containing a small number of more terms with derivatives.
As mentioned, this is one of the instances where things become complicated in the pure spinor formalism.

The $b$ operator in $D=10$ was constructed in ref. \cite{Berkovits:2001rb} using non-minimal variables, and reads
\begin{align}
b&=b_0+b_1+b_2+b_3\nn\\
&=-\fr2(\ll\lb)^{-1}(\lb\g^aD)\*_a\nn\\
&+\fr{16}(\ll\lb)^{-2}(\lb\g^{abc}d\lb)\left[N_{ab}\*_c
                         -\fr{24}(D\g_{abc}D)\right]\\
&+\fr{64}(\ll\lb)^{-3}(d\lb\g^{abc}d\lb)(\lb\g_aD)N_{bc}\nn\\
&-\fr{1024}(\ll\lb)^{-4}(\lb\g^{ab}{}_id\lb)(d\lb\g^{cdi}d\lb)N_{ab}N_{cd}\;,\nn
\end{align}
where $N=(\lambda\omega)$ and $N_{ab}=(\lambda\gamma_{ab}\omega)$ are invariant operators, in the sense that they respect the pure spinor constraint $(\lambda\gamma^a\lambda)=0$. 

The $b$ operator in $D=11$ \cite{Cederwall:2012es,Berkovits:2017xst} is somewhat more complicated. We will not display the full expression, but note that it is singular on the $16$-dimensional subspace (like the negative ghost number operators encountered in Section \ref{SGSection}), and begins as
\begin{align}
b=((\lambda\gamma_{de}\lambda)(\bar\lambda\gamma^{de}\bar\lambda))^{-1}
(\bar\lambda\gamma_{ab}\bar\lambda)(\lambda\gamma^{ab}\gamma^cD)\*_c+\ldots
\end{align}

There is also a possibility to find a $b$ operator that acts within functions of the minimal variables 
\cite{CederwallBOperatorUnpublished}, using the principles of ref. \cite{Cederwall:2010zz}.
On (holomorphic) functions of a $D=10$ pure spinor $\lambda$, the `ìnvariant derivative operator''
\begin{align}
\tilde\omega_\alpha=\omega_\alpha-\fr{4(N+3)}(\gamma^a\lambda)_\alpha(\omega\gamma_a\omega)
\end{align}
acts exactly like $\omega_\alpha$ between monomials $\lambda^{\alpha_1}\ldots\lambda^{\alpha_p}$, and annihilates the ideal generated by $(\lambda\gamma^a\lambda)$.
The minimal $b'$ operator reads
\begin{align}
b'=\fr{(N+4)(N+5)(N+6)}\left[\fr2(N^2+9N+15)(\tilde\omega\gamma^aD)\*_a
-\fr{128}N^{ab}(\tilde\omega D^3_{ab})\right]\;,\label{BPrimeEq}
\end{align}
where $D^3_{ab}$ is the antisymmetric product of three $D$'s in
$(01001)$,
\begin{align}
(D^3)_{ab}^\alpha=(\gamma^i)^{\alpha[\beta}(\gamma_{abi})^{\gamma\delta]}D_\beta D_\gamma D_\delta\;.
\end{align}
It can be shown explicitly that $b$ and $b'$ differ by a $Q$-exact expression.
Using $b'$, it is seen directly that Siegel gauge implies Lorenz gauge for the Yang--Mills connection. Namely, acting on
a ghost number $0$ field $\Psi=\lambda^\alpha A_\alpha$,
$b'\Psi=\fr{16}\*^a(D\gamma_a A)$.
A similar minimal $b$ operator should exist for any supersymmetric theory with local symmetries, \eg\ $D=11$ supergravity, but has not been constructed.

It may rightly be claimed that gauge fixing, in the form presented here, is rudimentary,
and more or less implemented at a first-quantised level.
A proper field-theoretic BV gauge fixing
\cite{Henneaux:1985kr}, involving a gauge fixing fermion, has not been developed in pure spinor superfield theory.

\subsection{Perturbative results}

The construction sketched above gives a recipe for calculating scattering amplitudes in the pure spinor formalism.
Any diagram --- which will contain a large number of component field diagrams --- should be saturated with appropriate vertex operators \cite{Aisaka:2009yp} representing external states.
There is a remaining issue of regularisation at $\lambda=0$ which was addressed and solved in ref. 
 \cite{Berkovits:2006vi}. This is due to the $b$ operators in propagators containing negative powers of $(\lambda\bar\lambda)$ that ultimately risk to make integrals divergent. 
Explicit evaluation of the regulated integrals is in general extremely complicated.
Results exist for superstring theory on Minkowski space \cite{Berkovits:2004dt,Berkovits:2006bk,Berkovits:2006vi,Aisaka:2009yp,Mafra:2011nv}
and on anti-de Sitter space \cite{Berkovits:2000yr,Berkovits:2004xu}.

The degree of convergence of loop diagrams in pure spinor superfield theory is generically much better than for loop diagrams in a component formulation or with superfields manifesting some fraction of supersymmetry.
Typical behaviour is the vanishing of bubbles and triangles in off-shell diagrams, \ie, as subdiagrams of any diagram.
For maximal super-Yang--Mills theory in $D=4$ \cite{Bjornsson:2010wu,Ben-Shahar:2021doh}, power counting is enough to demonstrate perturbative finiteness. 
In maximal supergravity in $D=4$ \cite{Cederwall:2012es,Karlsson:2014xva,Grassi:2011uf,Anguelova:2004pg}, power counting shows finiteness up to $6$ loops, and possibly a divergence at $7$ loops, see also refs.
\cite{Vanhove:2010nf,Bern:2009kd,Bern:2018jmv}. The precise statement is that an $L$-loop diagram is convergent in $D$ dimensions if $D<2+{14\over L}$, while for super-Yang--Mills theory it reads 
$D<4+{6\over L}$.

\section{Remarks\label{RemarksSection}}

Some final remarks, concerning shortcomings of the present status of pure spinor superfield theory, and some desirable developments. 

The classical theory of pure spinor superfields exhibits a striking simplicity.
Quantum calculations tend to become cumbersome, although in principle well defined, mainly due to the complicated expression for the $b$ operator used in gauge fixing, and the regularisation it brings along.
It remains an open question if these calculations can be simplified, either by finding a replacement for the $b$ operator, or by some completely different means. 

One approach, which has not been properly explored, would be to use the minimal (holomorphic) version of the $b$ operator,
$b'$ of eq. \eqref{BPrimeEq}. The construction will certainly extend to other negative ghost number operators
\cite{Cederwall:2010zz,CederwallBOperatorUnpublished}.
Possibly, in such a framework, the r\^ole of the non-minimal variables can be limited to integration, with the 
``simple'' regularisation of eq. \eqref{SimpleRegEq},
and the complicated regularisations at $\lambda=0$ may be avoided.

An urgent question for supergravity is the lack of manifest background invariance of the action
\eqref{FullSGActionEQ}.
 This is of course usual in string theory and string field theory
\cite{Zwiebach:1992ie} (see however ref. \cite{Sen:1994kx}), but one should be able to do better in a supergravity theory.
Indeed, even if the basis of the construction is in supergeometry, the geometric picture is lost in the final form.
There is some hope for ``re-geometrisation'', and for an understanding how to deform the model to non-flat backgrounds. 
It relies on deforming the algebra which is Koszul dual to functions of 
$11$-dimensional pure spinors \cite{CederwallJonssonPalmkvistSaberi},
in a manner similar to ref.
\cite{Figueroa-OFarrill:2015tan}.

As mentioned in Section \ref{GaugeFixingSection}, a proper field-theoretic BV gauge fixing procedure has not been developed for pure spinor superfield theory. There is no doubt that this can be done. It is probably one of the most important points on which the framework should be developed.

The whole idea about the formalism presented is to manifest as much symmetry as possible.
It is well known that dimensional reductions of $D=11$ enjoys U-duality, and that this symmetry 
can be ``geometrised'' within the context of 
exceptional geometry 
\cite{Cederwall:2017fjm,Cederwall:2019bai,Hohm:2019wql,Butter:2018bkl}.
Can the pure spinor framework be extended to accomodate for these symmetries?
Such a task may be very difficult, due to the infinite reducibility of local symmetries in extended geometry, since the pure spinor superfields always are based on the lowest-dimensional ghost field.
Indeed, already double supergeometry \cite{Cederwall:2016ukd,Butter:2022gbc} contains infinite reducibility in the Ramond-Ramond sector.

\newpage

\bibliographystyle{utphysmod2}

\phantomsection

\addcontentsline{toc}{section}{References}


\providecommand{\href}[2]{#2}\begingroup\raggedright\endgroup

\end{document}